\documentclass[aps, pre, onecolumn, longbibliography]{revtex4-2}
\usepackage{amssymb, amsfonts, amsmath, amsthm, mathtools, mathrsfs}
\usepackage[usenames, dvipsnames]{xcolor}
\usepackage[colorlinks, linkcolor=blue, anchorcolor=red, citecolor=blue]{hyperref}
\usepackage{braket}
\usepackage{orcidlink}
\usepackage[T1]{fontenc}
\usepackage[utf8]{inputenc}
\usepackage{lmodern}
\usepackage[english]{babel}

\begin{document}

\title{Large Deviation Functions of Open Quantum Systems with Strong Symmetries}

\author{Jiayin Gu\textsuperscript{1}\,\orcidlink{0000-0002-9868-8186}}
\author{Hailong Wang\textsuperscript{2}} 
\author{Shanhe Su\textsuperscript{3}}
\author{Fei Liu\textsuperscript{2}\,\orcidlink{0000-0002-4396-2977}}\email{feiliu@buaa.edu.cn}
\affiliation{\textsuperscript{\rm 1}School of Physics and Technology, Nanjing Normal University, Nanjing 210023, China}
\affiliation{\textsuperscript{\rm 2}School of Physics, Beihang University, Beijing 100083, China}
\affiliation{\textsuperscript{\rm 3}Department of Physics, Xiamen University, Xiamen 361005, China}

\date{\today}

\begin{abstract}
In open quantum systems with strong symmetries, the global rate function is typically non-convex and thus cannot be directly computed from the global scaled cumulant generating function (SCGF) via the G\"artner-Ellis theorem. To overcome this limitation, we develop a framework that applies the theorem locally within each symmetry sector: local rate functions are computed for each sector, and the global rate function is subsequently reconstructed as the minimum envelope of these local functions. We illustrate our method using an exactly solvable model and an open Heisenberg chain with XXX interactions. For the latter system with continuous non-Abelian symmetries, our approach reveals sector-dependent fluctuations ranging from Poissonian to super‑Poissonian statistics, together with emergent intermittency, demonstrating the broad applicability of our framework.

\end{abstract}

\maketitle

{\it Introduction}. -- Large deviation theory provides a general mathematical framework for characterizing nonequilibrium fluctuations~\cite{Ellis_2006, Touchette_PhysRep_2009, Vulpiani_2014, Burenev_SciPostPhysLectNotes_2025}. Within this formalism, the rate function and the scaled cumulant generating function (SCGF), which quantify rare fluctuation events, occupy a central role. These two functions are connected by the celebrated G\"artner-Ellis theorem: when the SCGF is differentiable, the rate function is given by the Legendre-Fenchel transform of the SCGF. The SCGF can be computed via the spectral methods~\cite{Lebowitz_JStatPhys_1999, Esposito_RevModPhys_2009} or simulation-based approaches~\cite{Giardina_PhysRevLett_2006, Lecomte_JStatMech_2007, Cavallaro_JPhysA_2016, Carollo_PhysRevE_2020, Bolhuis_AnnRevPhysChem_2002}. Evaluation of these two functions has underpinned substantial progress across a broad range of physical systems~\cite{Garrahan_PhysRevLett_2007, Harris_JStatMech_2007, Mehl_PhysRevE_2008, Prados_PhysRevLett_2011, PerezEspigares_PhysRevE_2013, Znidaric_PhysRevLett_2014, Carollo_PhysRevE_2020, Jack_EurPhysJB_2020, Liu_PhysRevE_2024, Liu_PhysRevE_2026}.

For open quantum systems with a unique stationary state~\cite{Evans_CommunMathPhys_1977, Albert_PhysRevA_2014}, the rate function is strictly convex~\cite{Esposito_RevModPhys_2009, Garrahan_PhysRevLett_2010, Schaller_2014, Rudge_JChemPhys_2019, Landi_PRXQuantum_2024}. However, this property breaks down for open quantum systems with strong symmetries, where ergodicity is broken and multiple degenerate stationary states emerge~\cite{Buca_NewJPhys_2012, Zhang_JPhysA_2020}. Manzano and collaborators~\cite{Manzano_PhysRevB_2014, Manzano_AdvPhys_2018, Manzano_NewJPhys_2021}, along with Munoz et al.~\cite{Munoz_PhysRevA_2019}, have shown that strong symmetries make the global SCGF non-analytical at zero counting field. Hence, the Gärtner‑Ellis theorem, which requires differentiability, no longer applies. Indeed, the Legendre-Fenchel transform of a non-anlytic SCGF would yield a convex envelope instead of the true rate function~\cite{Touchette_PhysRep_2009}. Since rate functions are directly accessible through experimental measurement, resolving this discrepancy is essential for the physical interpretation of observed nonequilibrium fluctuations. 

Despite this global limitation, open quantum systems with strong symmetries possess a well-established structural feature: their Hilbert spaces decompose into invariant subspaces, or symmetry sectors. These sectors are classified by system symmetries, and each supports at least one stationary state~\cite{Buca_NewJPhys_2012, Manzano_NewJPhys_2021, Zhang_JPhysA_2020}. Inspired by this sectorized structure, we conjecture an  intuitive physical picture for constructing the true rate function. Dynamical fluctuations take place within individual sectors, giving rise to {\it local} rate functions. In the large-time limit, the {\it global} rate function, which describes fluctuations over the full ensemble of trajectories, emerges from competition among these local rate functions. Since individual symmetry sectors cannot be decomposed further, each local rate function is expected to be strictly convex, so the G\"artner-Ellis theorem becomes applicable within each sector. Accordingly, each local function can be reliably computed from its corresponding {\it local} SCGF via the Legendre-Fenchel transform. In this paper, we implement this physical picture and verify it using two concrete open quantum systems.

{\it Strong Symmetries}. -- Let us consider an open quantum system, for which the Hilbert space is denoted by ${\cal H}$ and the space of bounded linear operators by ${\cal B}({\cal H})=({\cal H},{\cal H})$. The reduced density matrix $\rho(t)\in{\cal B}({\cal H})$ of the system satisfies the Markovian quantum master equation $\partial_t\rho(t)={\cal L}\rho(t)$, which explicitly reads
\begin{align}
\partial_t\rho=-\text{i}[H,\rho]+\sum_{i=1}^N \left(L_i\rho{L_i}^\dag-\frac{1}{2}\{{L_i}^\dag L_i,\rho\}\right) \text{,} \label{master_equation}
\end{align}
where the reduced Planck constant $\hbar$ is set to 1, $H$ represents the system's Hamiltonian, $L_i$ for $i=1,\cdots,N$ are the set of jump operators~\cite{Davies_CommunMathPhys_1974, Lindblad_CommunMathPhys_1976, Gorini_JMathPhys_1976}. We assume that the system possesses a {\it complete} set of strong symmetries $\{S_1,S_2,\cdots,S_n\}$, meaning that their matrix representations $R(S_k)$ commute with the Hamiltonian and all the jump operators~\cite{Buca_NewJPhys_2012}. These symmetries form a group and the group is allowed to be non-Abelian, i.e., $[R(S_k),R(S_{k'})] \neq 0$ for $k\neq k'$~\cite{Zhang_JPhysA_2020}. The group representation $R$ can be constructed via a direct sum of irreducible representations, $R=\oplus_{\alpha}m_{\alpha}R_{\alpha}$, where $m_{\alpha}$ denotes the multiplicity of the irreducible representation $R_\alpha$. The sector for each representation is denoted by ${\cal H}_{\alpha}$. In this construction, we require that ${\cal H}=\oplus_{\alpha}{\cal H}_{\alpha}$ and  $\dim({\cal H}_{\alpha})=m_{\alpha}\dim(R_{\alpha})$. In each sector, the Hamiltonian and jump operators are reduced to smaller-dimensional forms: $H_{\alpha}=\braket{{\cal H}_{\alpha}|H|{\cal H}_{\alpha}}$ and $L_{i,\alpha}=\braket{{\cal H}_{\alpha}|L_i|{\cal H}_{\alpha}}$. Moreover, the symmetries also induce a decomposition of the system's operator space ${\cal B}({\cal H})$ into a direct sum of blocks ${\cal B}_{\alpha\beta}=({\cal H}_{\alpha},{\cal H}_{\beta})$. Previous studies have shown that, under strong symmetries, the generator ${\cal L}$ of the quantum master equation~(\ref{master_equation}) cannot move an operator between different blocks~\cite{Buca_NewJPhys_2012}. Formally, this is equivalent to stating that the generator is block-diagonal in the operator space, i.e., ${\cal L}=\oplus_{\alpha}\oplus_{\beta}{\cal L}_{\alpha\beta}$, and the action of the reduced generator ${\cal L}_{\alpha\beta}$ on an operator $\rho_{\alpha\beta}\in{\cal B}_{\alpha\beta}$ remains within the same block ${\cal B}_{\alpha\beta}$. The relevant reduced generators for this work are ${\cal L}_{\alpha\alpha}$. Their formulas are obtained by replacing $H$, $L_i$, and $L^\dag_i$ in Eq.~(\ref{master_equation}) with the reduced $H_\alpha$, $L_{i,\alpha}$, and $L^\dag_{i,\alpha}$, respectively. Zhang et al. have demonstrated that in each diagonal block ${\cal B}_{\alpha\alpha}$, there exists one stationary state with degeneracy $\dim(R_{\alpha})^2$~\cite{Zhang_JPhysA_2020}. Thus, if the system's initial state lies within one of the diagonal blocks, the dynamics remain confined to that block, and the system approaches the stationary state in the large-time limit.

{\it Large Deviation Functions}. -- Equation~(\ref{master_equation}) can be unraveled into quantum jump trajectories~\cite{Carmichael_PhysRevA_1989, Dalibard_PhysRevLett_1992, Dum_PhysRevA_1992b, Wiseman_PhysRevA_1993,Plenio_RevModPhys_1998}. These trajectories consist of alternating deterministic pieces and random jumps of the wave functions of individual quantum systems~\cite{Breuer_2002}. Consider a quantum jump trajectory with $n$ jumps. Let the instant and type of each jump be $t_k$ and $a_k$, respectively, for $k=1,\cdots, n$. We denote this trajectory as $X_n(t)=(a_1,a_2,\cdots,a_n)$, where $t$ represents the duration of the process. The jump type $a_k$ is specified by the jump operators in Eq.~(\ref{master_equation}); therefore, it takes one of the values from $1$ to $N$. We define a counting variable for the trajectory $C[X_n(t)]=\sum_{k=1}^n w_{a_k}$, where the coefficient $w_{a_k}$ represents a weight assigned according to the jump type $a_k$ at the jump instant $t_k$. In this Letter, we focus on the case where all weights are simply equal to $1$. Thus, the counting variable represents the total activity $A$~\cite{Garrahan_PhysRevLett_2010}. Since the quantum jump trajectories are intrinsically random, the total activity $A$ follows a probability distribution. The large deviation principle states that in the large-time limit, the distribution $p(A,t)$ obeys the exponential approximation,
\begin{align}
p(A,t)\approx {\rm e}^{-tI(a)} \text{,} \label{probability_distribution}
\end{align}  
where  $a=A/t$ represents the activity rate and the exponent $I(a)$ is the rate function~\cite{Touchette_PhysRep_2009}.

It is generally very difficult, if not impossible, to directly obtain the distribution or the rate function. A convenient approach is to resort to the moment-generating function $Z(\epsilon,t)$ of $p(A,t)$ and its inverse transform. Notice that the moment-generating function is essentially the Laplace transform of the distribution with respect to the counting field $\epsilon$. Crucially, this function can be obtained by solving the tilted quantum master equation $\partial_t \tilde \rho(t)={\cal L}_{\epsilon}\tilde\rho (t)$ and taking the trace of $\tilde \rho(t)$, where the tilted generator   
\begin{align}
{\cal L}_{\epsilon}(\bullet)\equiv{\rm i}[\bullet,H]+\sum_{i=1}^N \left({\rm e}^{\epsilon}L_i\bullet{L_i}^\dag-\frac{1}{2}\{{L_i}^\dag L_i,\bullet\}\right) \text{,} \label{tilted_generator}
\end{align}
and the initial condition is the initial density matrix $\rho(0)$~\cite{Bagrets_PhysRevB_2003, Esposito_RevModPhys_2009,Garrahan_PhysRevLett_2010, Liu_PhysRevE_2016, Landi_PRXQuantum_2024}. Owing to the same strong symmetries, the tilted generator is also block-diagonal in the operator space, i.e., $
{\cal L}_\epsilon=\oplus_{\alpha}\oplus_{\beta}{\cal L}_{\epsilon,\alpha\beta}$. Thus, we obtain a formal solution for the moment-generating function as   
\begin{eqnarray}
\label{full_MGF}
Z(\epsilon,t)&=&{\rm Tr}\left[\oplus_{\alpha}\oplus_{\beta} \exp\left(t{\cal{L}}_{\epsilon,\alpha\beta}\right)\rho(0)\right]\nonumber \\
&=&\sum_{\alpha} {\rm Tr}\left[\exp\left(t{\cal{L}}_{\epsilon,\alpha\alpha}\right)\rho_{\alpha\alpha}(0)\right]\nonumber \\
&=&\sum_\alpha Z_\alpha(\epsilon,t),
\end{eqnarray}
where $\rho_{\alpha\alpha}(0)$ is the  projection of the initial density matrix $\rho(0)$ onto the diagonal block ${\cal B}_{\alpha\alpha}$, and ${\cal L}_{\epsilon,\alpha\alpha}$ is the reduced tilted generator, which is given by Eq.~(\ref{tilted_generator}) with the Hamiltonian and jump operators therein replaced by the reduced $H_\alpha$, $L_{i,\alpha}$, and $L^\dag_{i,\alpha}$, respectively. 
The second equality of Eq.~(\ref{full_MGF}) holds due to the orthogonality between distinct symmetry sectors. In the third equality, $Z_\alpha(\epsilon,t)$ is precisely the moment-generating function of a specific open quantum system, whose dynamics is governed by the reduced generator ${\cal L}_{\alpha\alpha}$, and whose quantum jump trajectories evolve within the sector ${\cal H}_{\alpha}$. Let the joint distribution of the counting variable for the constrained dynamics be $p_\alpha(A,t)$, which is by construction not normalized to unity. This is also obtained via the inverse transform of $Z_\alpha(\lambda,t)$. Then, performing the inverse transform on Eq.~(\ref{full_MGF}) yields 
\begin{eqnarray}
\label{full_counting_prob}
p(A,t)=\sum_\alpha p_\alpha(A,t). 
\end{eqnarray}

It is worth emphasizing that all symmetries of the open quantum system have been identified and the corresponding symmetry sectors cannot be further decomposed into smaller irreducible subspaces. Therefore, the local SCGFs in the diagonal blocks are differentiable, and the G\"artner-Ellis theorem is applicable within individual sectors. This indicates that, in the large-time limit, the local moment-generating function $Z_\alpha(\epsilon,t)$ and the local distribution $p_{\alpha}(A,t)$ can be approximated as $\exp[t\xi_{\alpha}(\epsilon)]$ and $\exp[-tI_{\alpha}(a)]$, respectively. In particular, the local rate function $I_\alpha(a)$ is related to the local SCGF $\xi_\alpha(\epsilon)$ via the Legendre-Fenchel transform:
\begin{align}
I_\alpha(a)=\sup_{\epsilon}\{a\epsilon-\xi_\alpha(\epsilon)\} \text{.} \label{Legendre_Fenchel}
\end{align}
Here, $\xi_\alpha(\epsilon)$ corresponds to the largest real part of the eigenvalues of the reduced tilted generator ${\cal L}_{\epsilon,\alpha\alpha}$~\cite{Bagrets_PhysRevB_2003, Esposito_RevModPhys_2009, Landi_PRXQuantum_2024}. Substituting these approximated counting distributions into Eq.~(\ref{full_counting_prob}) and taking the large-time limit, we obtain the global rate function as 
\begin{align}
I(a)=\min_{\alpha} \{I_{\alpha}(a)\} \text{,} \label{global_rate_function}
\end{align}
where the minimization is performed over all diagonal blocks that have a non-zero projection of the initial density matrix. Equation~(\ref{global_rate_function}) is the central result of this Letter. Clearly, the global rate function of open quantum systems with strong symmetries is generally non-convex. 

Before moving to concrete models, we explain the importance of firstly performing the inverse transform and then taking the large-time limit. If this order is reversed, in the large-time limit, substitution of the approximated local SCGFs into  Eq.~(\ref{full_MGF}) yields the global SCGF 
\begin{eqnarray}
\label{globalblockSCGFs}
\xi(\epsilon)= \max_\alpha\{ \xi_\alpha(\epsilon)\}. 
\end{eqnarray} 
Generally, $\xi_{\alpha}(\epsilon)$ differs for different $\alpha$. Thus, $\xi(\epsilon)$ is non-analytic at $\epsilon=0$~\cite{Manzano_PhysRevB_2014, Manzano_AdvPhys_2018, Manzano_NewJPhys_2021,Munoz_PhysRevA_2019}.
As already mentioned, applying the Legendre-Fenchel transform to this  non-analytic function would result in a convex envelope of the true global rate function. Our ordering is also supported by a dissipative freezing phenomenon revealed in open quantum systems with strong symmetries~\cite{Munoz_PhysRevA_2019, Tindall_SciPostPhysCore_2023, Halati_PhysRevRes_2022,Manzano_NewJPhys_2021}: Even for an initially prepared quantum superposition state, the wave function of the quantum jump trajectory will always become frozen within one of the sectors after sufficient evolution time, and the freezing probability in a certain sector is determined by the squared norm of the initial wave function within that sector. This implies that, after a finite period, two quantum jump trajectories, one with an initial wave function in a quantum superposition of different sectors, and the other with an initial wave function confined to a single sector, will become identical. In the large-time limit, Eq~(\ref{full_counting_prob}) emerges as a natural consequence of this quantum phenomenon.

{\it An Analytical Model}. -- In this model, both the Hamiltonian and the only jump operator are assumed to be proportional to a symmetry operator $S$, e.g., the $x$-component of a collective spin operator in Ref.~\cite{Munoz_PhysRevA_2019}. Because there exists only one symmetry, this system is Abelian and the irreducible index reduces to the eigenvalue of the symmetry operator. Let the non-degenerate eigenvalue and eigenstate of $S$ be $m$ and $\ket{m}$ for $m=1,\cdots$. Then, the sector ${\cal H}_m$ is $\{\ket{m}\}$, and the eigenvalue of the reduced tilted generator ${\cal L}_{\epsilon,m}$, i.e., the local SCGF, is simply~\cite{Munoz_PhysRevA_2019},  
\begin{align}
\xi_{m}(\epsilon)= m^2({\rm e}^{\epsilon}-1) \text{.} \label{SCGF_first_model}
\end{align}
Obviously, as a function of $\epsilon$, the SCGFs are degenerate at $\epsilon=0$. Taking their Legendre-Fenchel transform with respect to $\epsilon$, we obtain the local rate functions for each sector as
\begin{align}
I_m(a)=a\ln\left(\frac{a}{m^2}\right)-a+m^2 \text{.} \label{rate_function_first_model}
\end{align}
Equation~(\ref{rate_function_first_model}) indicates that, if we observe a quantum jump trajectory for a long time, we will find that its activity statistics is always Poissonian with mean counting rate $m^2$. Using Eq.~(\ref{global_rate_function}), the global rate function $I(a)$ is obtained by taking the minimum among $\{I_m(a)\}$, given that the system's initial state has a non-zero component in the sector ${\cal H}_m$. For instance, if the system starts from the quantum superposition wave function $c_1\ket{1}+c_3\ket{3}$, then $I(a)=\min\{I_1(a),I_3(a)\}$. Our result agrees with Eq.~(23) given in Ref.~\cite{Munoz_PhysRevA_2019}, where the counting distributions have been solved exactly for each sector.

\begin{figure*}
\begin{minipage}[t]{0.45\hsize}
\resizebox{1.0\hsize}{!}{\includegraphics{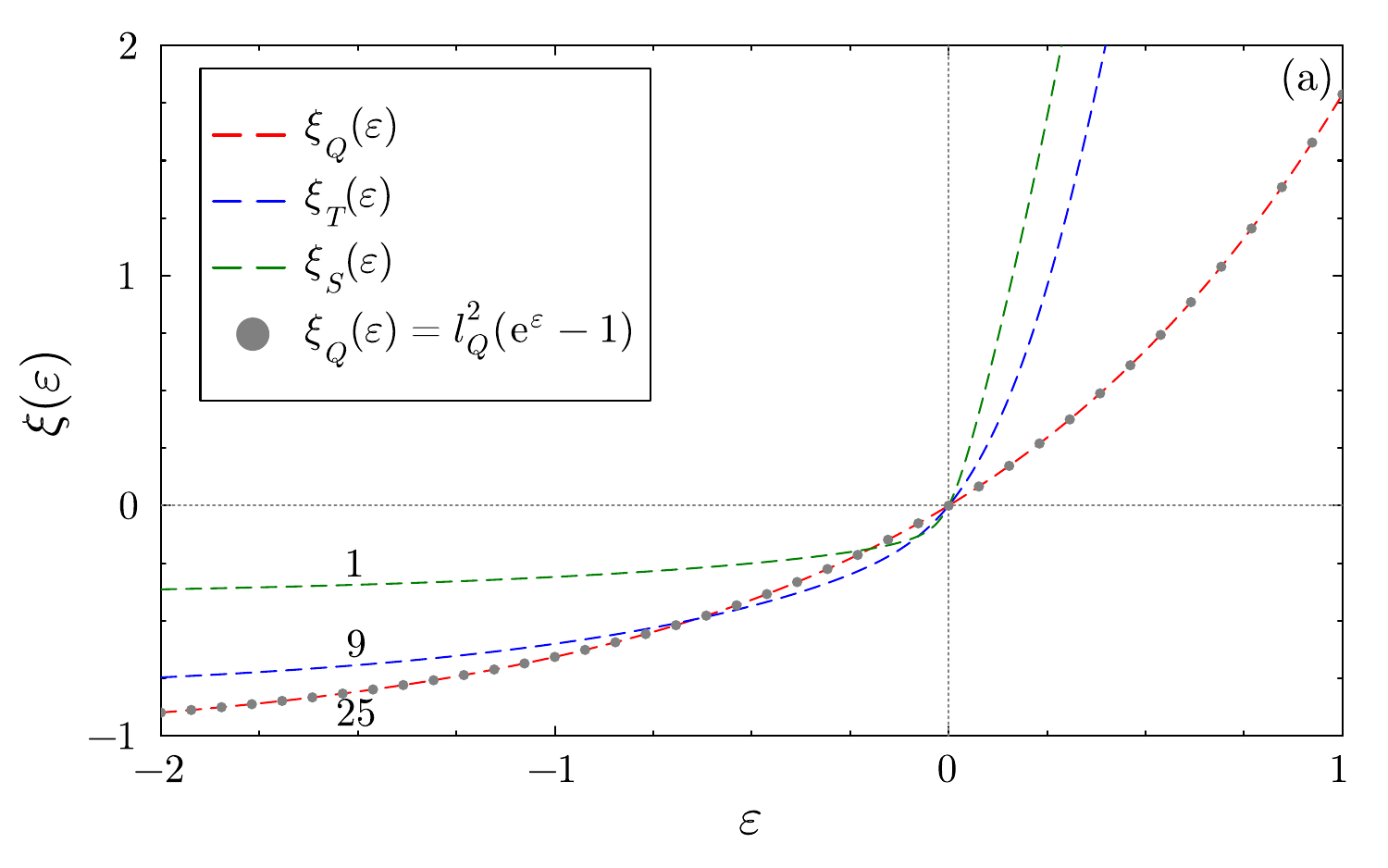}}
\end{minipage}
\begin{minipage}[t]{0.45\hsize}
\resizebox{1.0\hsize}{!}{\includegraphics{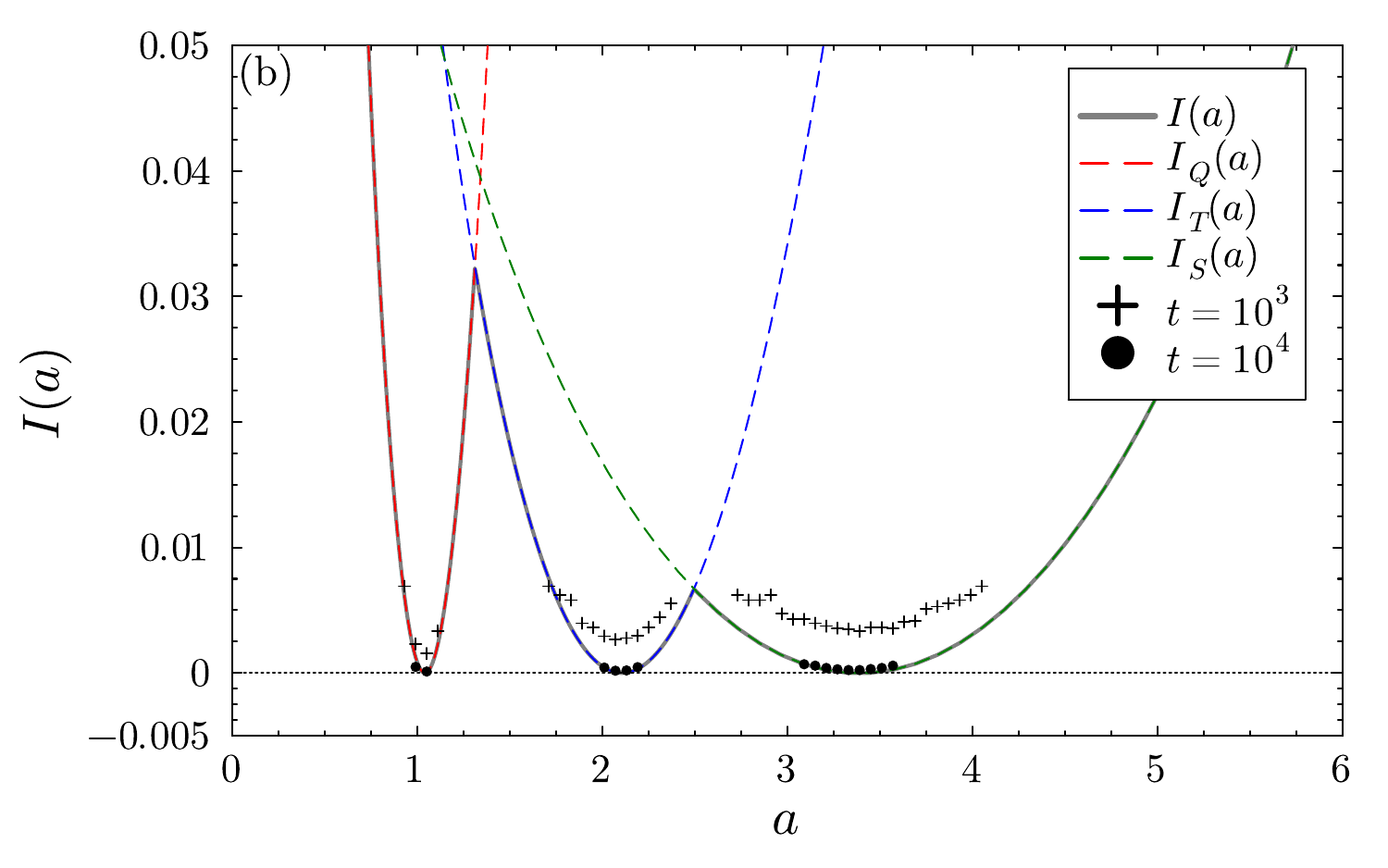}}
\end{minipage}
\begin{minipage}[t]{0.45\hsize}
\resizebox{1.0\hsize}{!}{\includegraphics{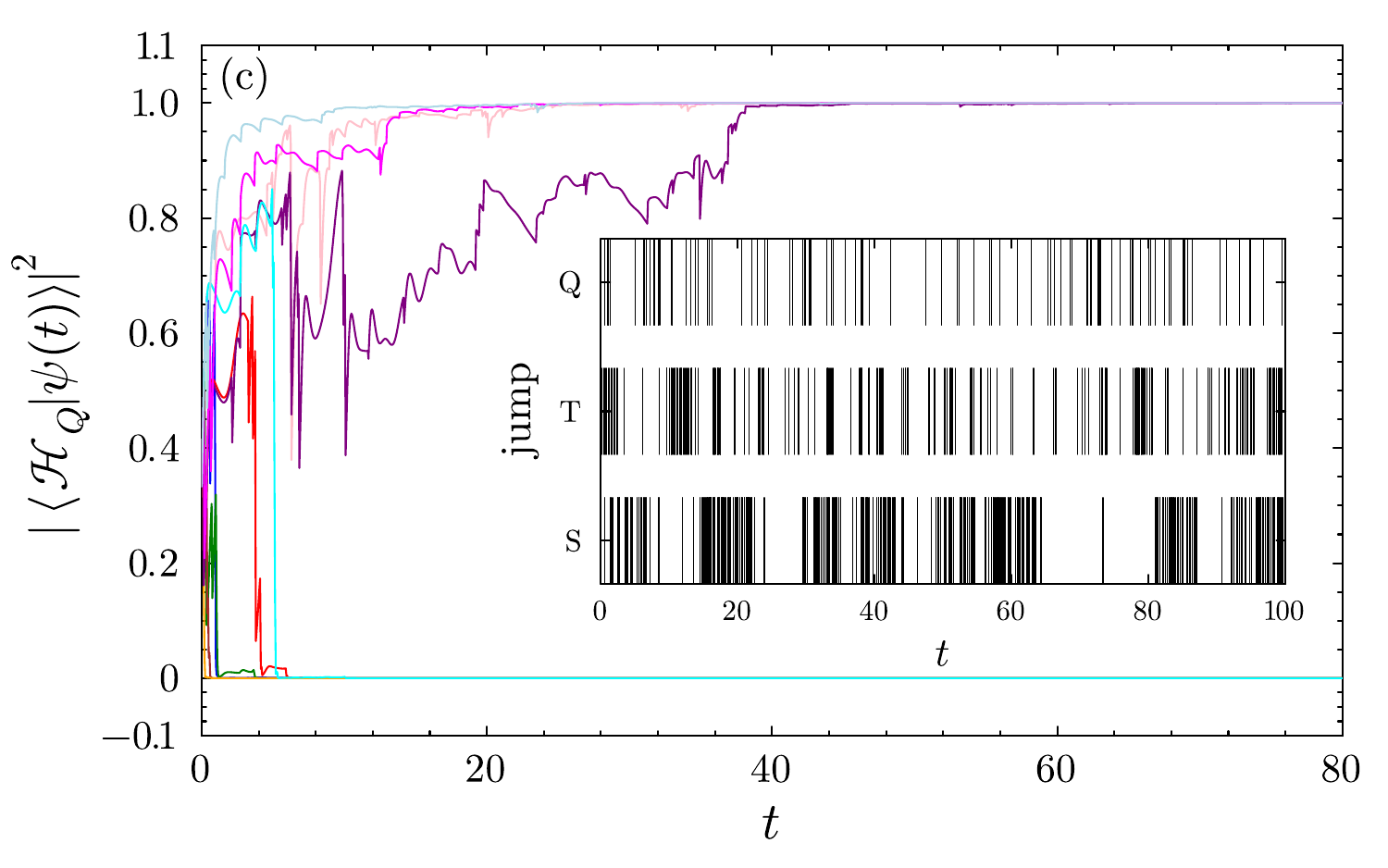}}
\end{minipage}
\begin{minipage}[t]{0.45\hsize}
\resizebox{1.0\hsize}{!}{\includegraphics{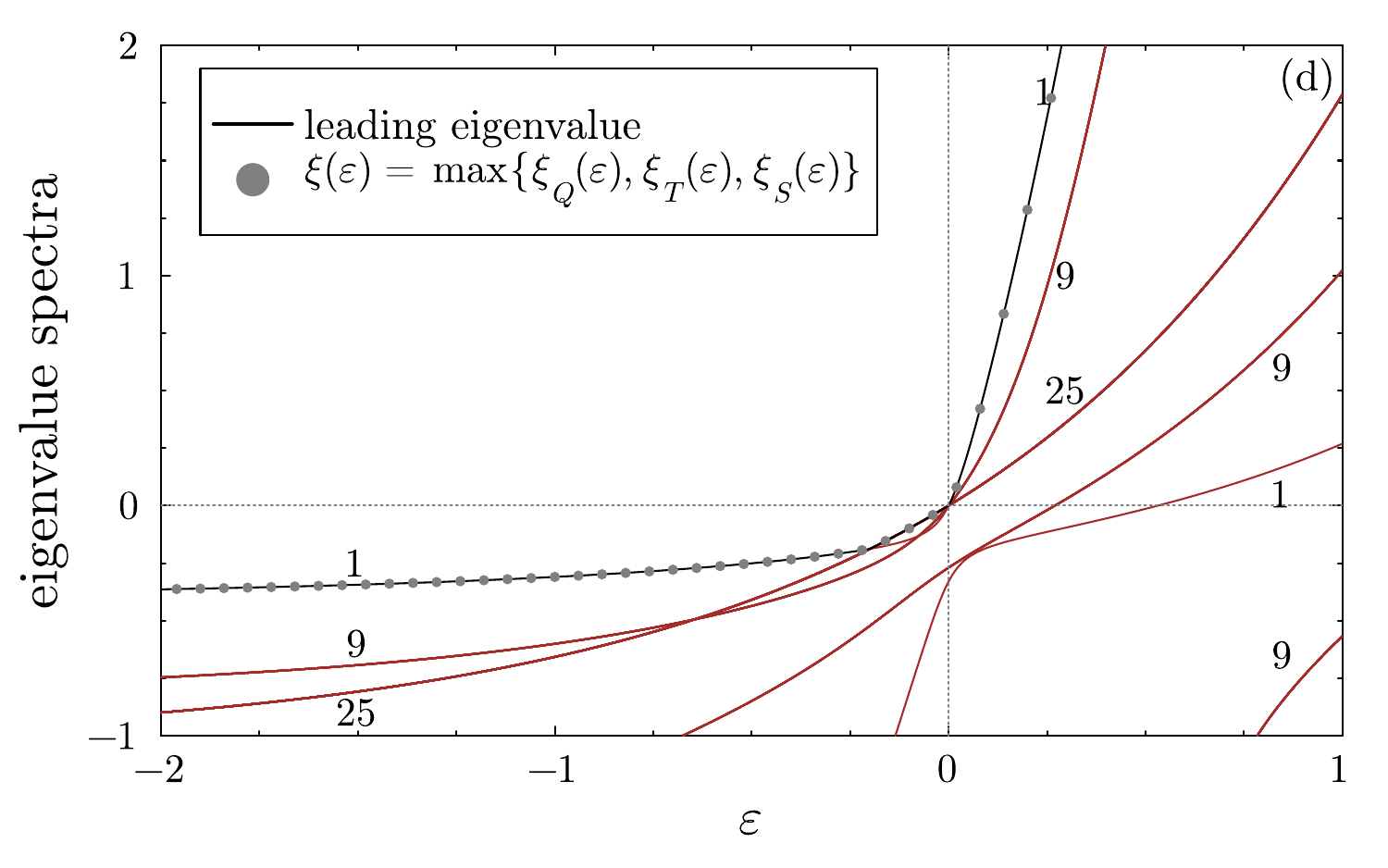}}
\end{minipage}
\caption{The results for the dissipative Heisenberg chain with four spins. (a) The three colored curves represent the local SCGFs in the three sectors obtained by numerically finding the real leading eigenvalue of the reduced tilted generators ${\cal L}_{Q,\epsilon}$, ${\cal L}_{T,\epsilon}$, and ${\cal L}_{S,\epsilon}$. The gray circles are the theoretical predictions of the SCGF in quintet sector. The numbers near the curves represent the multiplicities of the corresponding eigenvalue. (b) The three colored curves are the local rate functions obtained by applying the Legendre-Fenchel transform to their corresponding local SCGFs. The solid segment is the lower envelope of the three curves, making up the global rate function, which is non-convex. The crosses and circles denote the global rate function obtained by simulating $1000$ trajectories. The whole time intervals for the trajectories are $t=10^3$ and $10^4$, respectively. (c) Time evolution of the projection onto the sector ${\cal H}_Q$ spanned by quintet states for $10$ quantum jump trajectories. The inset shows three quantum jump trajectory segments, Q, T, S, each simulated and recorded within the quintet, triplet, and singlet sectors, respectively. (d) The 54 real leading eigenvalues (imaginary part smaller than $1.0\times 10^{-13}$) of full tilted generator ${\cal L}_{\epsilon}$. The numbers near solid lines represent the multiplicities of the corresponding eigenvalue. The black line is the leading eigenvalue that represents the global SCGF $\xi(\epsilon)$. The gray circles are the maxima of three local SCGFs in the sectors. For the results in all panels, the coefficients specifying the jump operator are $K_{1,2}=0.06$, $K_{2,3}=0.18$, and $K_{3,4}=0.78$. Thus, we have $l_Q=K_{1,2}+K_{2,3}+K_{3,4}=1.02$. In panels (b) and (c), the trajectories are simulated from the same initial superposition state $\psi(0)=(\ket{2,0}+\ket{1,0}+\ket{0,0})/\sqrt{3}$, where $\ket{2,0}=\left(\ket{\downarrow\downarrow\uparrow\uparrow}+\ket{\downarrow\uparrow\downarrow\uparrow}+\ket{\downarrow\uparrow\uparrow\downarrow}+\ket{\uparrow\downarrow\downarrow\uparrow}+\ket{\uparrow\downarrow\uparrow\downarrow}+\ket{\uparrow\uparrow\downarrow\downarrow}\right)/\sqrt{6}$, and $\ket{1,0}=\left(\ket{\uparrow\uparrow\downarrow\downarrow}-\ket{\downarrow\downarrow\uparrow\uparrow}\right)/\sqrt{2}$, and $\ket{0,0}=\left(\ket{\uparrow\downarrow\uparrow\downarrow}-\ket{\uparrow\downarrow\downarrow\uparrow}-\ket{\downarrow\uparrow\uparrow\downarrow}+\ket{\downarrow\uparrow\downarrow\uparrow}\right)/2$. Here, $\ket{S,M}$ denotes the state having total spin quantum number $S$, and total spin magnetic quantum number $M$. The tiny discretized time interval in simulating trajectories is ${\rm d}t=0.001$.}
\label{fig_res}
\end{figure*}

{\it Heisenberg chain with XXX Interactions}. -- The second model is an open isotropic XXX spin chain~\cite{Zhang_JPhysA_2020}. 
The Hamiltonian is 
\begin{align}
H=\sum_{i=1}^{N-1}\left(\sigma_i^x\sigma_{i+1}^x+\sigma_i^y\sigma_{i+1}^y+\sigma_i^z\sigma_{i+1}^z\right) \text{,}
\end{align}
where $\sigma_i^{x,y,z}$ are the three Pauli operators at the $i$-th site. The system is coupled to an environment that induces dissipation between adjacent sites with random strength, and the jump operator is set to 
\begin{align}
L=\sum_{i=1}^{N-1}K_{i,i+1}\left(\sigma_i^x\sigma_{i+1}^x+\sigma_i^y\sigma_{i+1}^y+\sigma_i^z\sigma_{i+1}^z\right) \text{.}
\end{align}
Because the Hamiltonian and jump operator commute with the total spin operators $S^{x,y,z}=(1/2)\sum_{i=1}^N\sigma_i^{x,y,z}$, 
\begin{align}
[H,S^{x,y,z}]=[L,S^{x,y,z}]=0 \text{,}
\end{align}
the open quantum system satisfies a continuous ${\rm SU}(2)$ strong symmetry. Owing to its non-Abelian characteristics, Zhang et al. have intensively studied the model with regard to the degeneracy of stationary states~\cite{Zhang_JPhysA_2020}. To be specific, we restrict to the system composed of four spin-$1/2$ particles. In this case, the group representation decomposes into ${\bf 2}^{\otimes 4}={\bf 5}\oplus(3\times{\bf 3})\oplus(2\times{\bf 1})$, where ${\bf 5}$ refers to the quintet representation, $3\times{\bf 3}$ the triplet representation with three multiplicities, and $2\times{\bf 1}$ the singlet representation of two multiplicities. Accordingly, we have sectors ${\cal H}_Q$ of dimension $5$, ${\cal H}_T$ of dimension $9$, and ${\cal H}_S$ of dimension $2$. The detailed quantum basis and matrix representations of the Hamiltonian and jump operator under this decomposition are given in Appendix A.
Generalization to a system with an arbitrary number of spins is straightforward.

According to Schur's lemma~\cite{Zee_2016}, the reduced Hamiltonian and the reduced jump operator in the $S=2$ sector are both multiples of the identity matrix, namely $H_Q=h_QI$ and $L_Q=l_QI$. Then, the SCGF is given by Eq.~(\ref{SCGF_first_model}) with $m$ therein replaced by $l_Q$, where $l_Q=K_{1,2}+K_{2,3}+K_{3,4}$~\footnote{In the quintet, the four spins are aligned in the same direction so that the total spin quantum number is $S_{\rm tot}=2$. The jump operator can be expressed as $L=4\sum_{i=1}^3K_{i,i+1}{\bf S}_i\cdot{\bf S}_{i+1}=4\sum_{i=1}^3K_{i,i+1}S_iS_{i+1}=K_{1,2}+K_{2,3}+K_{3,4}$, where we have used $S_i=S_{i+1}=1/2$.}. In the $S=1$ and $S=0$  sectors, the group representations have multiplicities greater than one. Thus, the two reduced operators are not simply multiples of identity operator. We have to numerically find the corresponding $\xi_T(\epsilon)$ and $\xi_S(\epsilon)$.
All three SCGFs are plotted in panel (a) of Fig.~\ref{fig_res}. Based on these data, the rate functions for the three sectors can be calculated through the Legendre-Fenchel transform~(\ref{Legendre_Fenchel}), and the results are presented in panel (b) of Fig.~\ref{fig_res}. Lastly, according to Eq.~(\ref{global_rate_function}), the global rate function is the lower envelope of the three curves, $I(a)=\min\{I_Q(a),I_T(a),I_S(a)\}$. We see that it is non-convex. To further verify this three-peak structure of the global rate function, we compare it with the data from direct sampling by simulating quantum jump trajectories~\cite{Plenio_RevModPhys_1998, Breuer_2002}. These trajectories are simulated from the same initial state that is superposed among the three sectors. As seen from the panel, when the time interval for the trajectories is as large as $10^4$, the global rate function obtained from the direct sampling converges well to the expected one, showing that the two independent methods are in excellent agreement with each other. This simulation also reveals the dissipative freezing phenomenon of the open Heisenberg chain. As shown in panel (c) of Fig.~\ref{fig_res}, after a period of relaxation, some trajectories are trapped in the quintet or $S=2$ sector, while others escape from it completely.

Several comments are in order. Firstly, the rate functions in panel (b) of Fig.~\ref{fig_res} indicate that the system's activity is distinct across the three sectors: Poissonian in the $S=1$ sector, and super-Poissonian in the other two sectors, respectively. This feature, where the variance is much larger than the mean for the latter two, also appears in the local SCGFs $\xi_T(\epsilon)$ and $\xi_S(\epsilon)$ in panel (a). Furthermore, these two curves, particularly that for the $S=0$ sector, exhibit rapid phase crossovers near the origin. Hence, the intermittency phenomenon is expected in their quantum jump trajectories~\cite{Plenio_RevModPhys_1998,Garrahan_PhysRevLett_2010}. This is demonstrated by the inset of panel (c) in Fig.~\ref{fig_res}. Secondly, a simple relationship exists between the local SCGFs and the non-analytical global SCGF. In quantum mechanics, decomposing the full Hilbert space into invariant subspaces essentially aims to obtain a block-diagonal Hamiltonian~\cite{Zee_2016}. In the case of open quantum systems, this corresponds to constructing a block-diagonal full tilted generator ${\cal L}_{\epsilon}$. Because the reduced tilted generators ${\cal L}_{\epsilon,\alpha\alpha}$ are exactly the diagonal blocks, their local SCGFs must appear among the real parts of the spectrum of the full tilted generator. This is verified in panel (d) of Fig.~\ref{fig_res}, where 54 leading real parts of the eigenvalues are plotted against the counting field $\epsilon$~\footnote{This feature provides an alternative approach for computing the local SCGFs: directly diagonalizing the full tilted generator and extracting all eigenvalue curves whose real parts pass through the origin. The advantage of this method is that it no longer resorts to group-theoretical techniques. However, its limitations are also substantial. First, the matrix of the tilted generator grows drastically with the system's degrees of freedom, rendering diagonalization computationally difficult. Second, the eigenvalue curves exhibit large degeneracies, so one must visually identify which data points belong to the smooth local SCGF within the same diagonal block.}. We see that the upper envelope of the three local SCGFs gives exactly the global SCGF, $\xi(\epsilon)=\max\{\xi_Q(\epsilon),\xi_T(\epsilon),\xi_S(\epsilon)\}$. Owning to the crossover of these lines, the global function becomes non-analytic at the origin. Finally, the multiplicities of the local SCGFs are equal to the square of the dimension of each irreducible representation. We specifically label these multiplicities near each line in panels (a) and (d). The degeneracy of stationary states is determined by $25+9+1=35$ zero eigenvalues at the origin~\cite{Zhang_JPhysA_2020}. An explanation is given in Appendix A. The inclusion of a non-zero counting field $\epsilon$ in the tilted generator~(\ref{tilted_generator}) does not break the system's symmetry, yet the degeneracy is no longer sustained among eigenvalues belonging to distinct diagonal blocks, as these blocks carry different activity rates.

{\it Conclusion}. -- This paper extends the applicability of the G\"artner-Ellis theorem to open quantum systems with strong symmetries and resolves the failure of global large-deviation analysis caused by non-analytic global SCGFs. Instead of performing global transformation in the full operator space, we apply the G\"artner-Ellis theorem within the diagonal blocks that are classified by the symmetries. After obtaining local rate functions from the local smooth SCGFs in individual blocks, the global rate function is then obtained as their
minimum envelope. We confirm the consistency of this framework using both an exactly solvable model and a dissipative Heisenberg spin chain, where quantum jump simulations support our theoretical results. In the near future, we will investigate practical open quantum systems within the present framework.

\smallskip

\textit{Acknowledgments}. -- This work was supported by the National Natural Science Foundation of China under Grant No. 12505048 and No. 12075016.

\appendix

\section{Multiplicities of the local SCGFs of dissipative Heisenberg chain} 
When combining four spin-$1/2$ particles of the dissipative Heisenberg chain, the total state space has a dimension of $2^4=16$ states. By using the rules of addition of angular momentum in quantum mechanics, these 16 states can be classified by their total spin quantum number $S$ and total magnetic quantum number $M$. Because there are multiple ways to couple the individual spins to arrive at the same total $S$, several representation-equivalent but distinct multiplets appear. Their number is known as the multiplicity. To write out the explicit basis in terms of the individual spin-up $\ket{\uparrow}$ and spin-down $\ket{\downarrow}$ configurations, we use standard angular momentum coupling, Clebsch-Gordan coefficients~\cite{Sakurai1994}. Let's group the basis by their total spin $S$ , total magnetic quantum number $M$, for $M=-S,\cdots,+S$, and the multiplicity index $\cal I$. The notation used is $\ket{S,M;{\cal I}}$. Then, the three sectors are 
\begin{align}
	& {\cal H}_Q={\rm span}\{\ket{2,+2},\ket{2,+1},\ket{2,0},\ket{2,-1},\ket{2,-2}\} \text{,} \\
	& {\cal H}_T={\rm span}\{\ket{1,+1;1},\ket{1,+1;2},\ket{1,+1;3},\ket{1,0;1},\ket{1,0;2},\ket{1,0;3},\ket{1,-1,1},\ket{1,-1,2},\ket{1,-1,3} \} \text{,} \\
	& {\cal H}_S={\rm span}\{\ket{0,0;1},\ket{0,0;2}\} \text{.}
\end{align}

According to Schur's lemma~\cite{Zee_2016}, the Hamiltonian and jump operator in the $S=2$ sector are multiples of the $5\times 5$ identity matrix, that is, $H_Q=h_QI_{5\times 5}$, and $L_Q=l_QI_{5\times 5}$. Substituting these results into Eq.~(\ref{eigenvlauesequation}), we obtain the eigenvalue $l_Q^2(\exp(\epsilon)-1)$. Moreover, for the same eigenvalue, there are distinct eigenmatrices, $|2,M\rangle\langle 2,M'|$, for $M,M'=-2,\cdots,+2$. Hence, the SCGF in the $S=2$ sector has a multiplicity $5^2=25$. The Hamiltonian within the $S=1$ sector is $H_T=I_{3\times 3}\otimes h_T$. Analogously, the jump operator is $L_T=I_{3\times 3}\otimes l_T$, where $h_T$ and $l_T$ are the representations of the Hamiltonian and the jump operator within the three-dimension multiplicity space, span$\{|1\rangle,|2\rangle,|3\rangle\}$, since $\langle 1,M;{\cal I} |H|1,M';{\cal I'}\rangle=0$ for $M\neq M'$. The Hamiltonian and jump operator within the $S=0$ sector are $2\times 2$ matrices $H_S=h_S$ and $L_S=l_S$, respectively, where $h_S$ and $l_S$ are the representations of the Hamiltonian and the jump operator within the two-dimension multiplicity space, span$\{|1\rangle,|2\rangle\}$. The reason is due to the unique magnetic quantum number, $M=0$. For convenience of the reader, we provide explicit expressions for the jump operators in the three- and two-dimensional multiplicity spaces:
\begin{align}
	&l_T=\begin{pmatrix}
		K_{1,2}+K_{3,4}-K_{2,3} & \sqrt{2}K_{2,3} &\sqrt{2}K_{2,3}  \\
		\sqrt{2}K_{2,3}  & K_{1,2}-3K_{3,4} & -K_{2,3} \\
		\sqrt{2}K_{2,3} &-K_{2,3} & -3K_{1,2}+K_{3,4} 
	\end{pmatrix} \text{,}\hspace{1cm}\\
 & l_S=\begin{pmatrix}
 	K_{1,2}+K_{3,4}-2K_{2,3} & \sqrt{3}K_{2,3} \\
 	\sqrt{3}K_{2,3}  & -3K_{1,2}-3K_{3,4}
 \end{pmatrix}  \text{.}
\end{align}
The Hamiltonians $h_T$ and $h_S$ are simply obtained by letting $K_{1,2}=K_{2,3}=K_{3,4}=1$ in these formulas, respectively.

As analyzed in the main text, the local SCGFs for the three sectors of the open quantum system are the real parts of leading eigenvalues of the reduced tilted generators. The latter can be solved by the eigenvalue equation given as 
   \begin{align}
   	{\cal L}_{\epsilon, \alpha}\rho^{(\alpha)}=\xi_\alpha(\epsilon)\rho^{(\alpha)} \text{,}  \label{eigenvlauesequation}
   \end{align}
for $\alpha=Q,T,S$. On the other hand, the eigenmatrix for $S=1$ sector can be formally written as  
\begin{align}
	\rho^{(T)}=\sum_{M,M'=-1}^{+1} (\ket{1,M}\!\bra{1,M'})\otimes \rho^{(T)}_{MM'} \text{,} \label{eigenmatrixdirectproductform}
\end{align}
where $\rho^{(T)}_{KK'}$ is the undetermined $3\times 3$ matrix, whose elements are ordered according to the multiplicity ${\cal I}=1,2,3$. Substituting $H_T$, $L_T$, and Eq.~(\ref{eigenmatrixdirectproductform}) into the left side of Eq.~(\ref{eigenvlauesequation}), we derive
\begin{align}
	& {\rm i}[\rho^{(T)},H_T]+{\rm e}^\epsilon L_T\rho^{(T)}L_T^\dag -\frac{1}{2}\{L_T^\dag L_T, \rho^{(T)}\}  \nonumber \\
	& \quad\quad\quad\quad\quad\quad =\sum_{M,M'=-1}^{+1}(\ket{1,M}\!\bra{1,M'})\otimes\left({\rm i}[\rho^{(T)}_{MM'},h_T]+{\rm e}^\epsilon l_T\rho^{(T)}_{MM'}l_T^\dag -\frac{1}{2}\{l_T^\dag l_T, \rho^{(T)}_{MM'}\}\right) \text{.} \label{lefthandsideofeigeneq}
\end{align}
Here the operator composition rule for tensor products, $(E\otimes F)(G\otimes H)=EG\otimes FH$, has been used. Obviously, the terms in the circle brackets of Eq.~(\ref{lefthandsideofeigeneq}) define a tilted generator with Hamiltonian $h_T$ and jump operator $l_T$. Substituting Eq.~(\ref{eigenmatrixdirectproductform}) into the right-hand side of Eq.~(\ref{eigenvlauesequation}) and letting it equal Eq.~(\ref{lefthandsideofeigeneq}), due to independence of different operators $\ket{1,M}\!\bra{1,M'}$, we arrive at
\begin{align}
	{\rm i}[\rho^{(T)}_{MM'},h_T]+{\rm e}^\epsilon l_T\rho^{(T)}_{MM'}l_T^\dag -\frac{1}{2}\{l_T^\dag l_T, \rho^{(T)}_{MM'}\}=\xi_T(\epsilon)\rho^{(T)}_{MM'} \text{,} \label{subeigenvlauesequation}
\end{align}
for $M,M'=-1,0,1$. Equation~(\ref{subeigenvlauesequation}) clearly indicates that the multiplicity of the local SCGF in the $S=1$ sector is $3^2=9$. It is worth pointing out that the computational effort of solving Eq.~(\ref{subeigenvlauesequation}) is much decreased compared to solving the original Eq.~(\ref{eigenvlauesequation}) for $
\alpha=T$, since the tilted generator of Eq.~(\ref{subeigenvlauesequation}) is a $9\times 9$ instead of $81\times 81$ matrix. Equation~(\ref{subeigenvlauesequation}) clearly shows that the fluctuation dynamics of the dissipative Heisenberg chain is governed by the Hamiltonian and jump operators acting on the three-dimensional multiplicity space, i.e., $h_T$ and $l_T$. Performing an analogous discussion, we will obtain the leading eigenvalue for the $S=0$ sector by solving a matrix with size $4\times 4$. In particular, its multiplicity is  $1^2=1$. 

Before closing this section, we wish to mention the relationship between the multiplicities of the local SCGFs and degeneracy of the system's stationary state~\cite{Zhang_JPhysA_2020}. Evidently, the stationary state is exactly the eigenmatrix of the full tilted generator ${\cal L}_{\epsilon=0}$ with zero eigenvalue. Hence, its degeneracy  is simply equal to $1+9+25=35$.

\end{document}